\documentclass[conference]{IEEEtran}
\IEEEoverridecommandlockouts
\usepackage{cite}
\usepackage{amsmath,amssymb,amsfonts}
\usepackage{algorithmic}
\usepackage{graphicx}
\usepackage{textcomp}
\usepackage{xcolor}
\usepackage[ruled, linesnumbered]{algorithm2e}

\usepackage{booktabs} % For formal tables

\usepackage{subfig}

\usepackage{multirow}

\usepackage{array}
\newcolumntype{L}[1]{>{\raggedright\let\newline\\\arraybackslash\hspace{0pt}}m{#1}}
\newcolumntype{C}[1]{>{\centering\let\newline\\\arraybackslash\hspace{0pt}}m{#1}}
\newcolumntype{R}[1]{>{\raggedleft\let\newline\\\arraybackslash\hspace{0pt}}m{#1}} 

\newcommand{\nop}[1]{}

\begin{document}
\title{Unsupervised User Identity Linkage via \\Factoid Embedding}
	
\author{\IEEEauthorblockN{Wei Xie\IEEEauthorrefmark{1}, Xin Mu\IEEEauthorrefmark{2}, Roy Ka-Wei Lee\IEEEauthorrefmark{1}, Feida Zhu\IEEEauthorrefmark{1} and Ee-Peng Lim\IEEEauthorrefmark{1}} 
	\IEEEauthorblockA{\IEEEauthorrefmark{1}Living Analytics Research Centre\\
		Singapore Management University, Singapore\\
		Email: \{weixie,roylee.2013,fdzhu,eplim\}@smu.edu.sg} 
	\IEEEauthorblockA{\IEEEauthorrefmark{2}National Key Laboratory for Novel Software Technology\\
		Nanjing University, China\\
		Email: mux@lamda.nju.edu.cn}
}

	\maketitle

	\begin{abstract}
		User identity linkage (UIL), the problem of matching user account across multiple online social networks (OSNs), is widely studied and important to many real-world applications. Most existing UIL solutions adopt a supervised or semi-supervised approach which generally suffer from \textit{scarcity of labeled data}. In this paper, we propose \textsf{Factoid Embedding}, a novel framework that adopts an unsupervised approach. It is designed to cope with different profile attributes, content types and network links of different OSNs. The key idea is that each piece of information about a user identity describes the real identity owner, and thus distinguishes the owner from other users. We represent such a piece of information by a \textit{factoid} and model it as a triplet consisting of \textit{user identity}, \textit{predicate}, and an \textit{object} or another \textit{user identity}. By embedding these factoids, we learn the user identity latent representations and link two user identities from different OSNs if they are close to each other in the user embedding space. Our \textsf{Factoid Embedding} algorithm is designed such that as we learn the embedding space, each embedded factoid is ``translated'' into a motion in the user embedding space to bring similar user identities closer, and different user identities further apart. Extensive experiments are conducted to evaluate \textsf{Factoid Embedding} on two real-world OSNs data sets. The experiment results show that \textsf{Factoid Embedding} outperforms the state-of-the-art methods even without training data. 
		
		%The key idea behind Factoid Embedding is that despite the heterogeneity in information from multiple OSNs, each piece of information about a user identity still describes the real person who owns it, and consequently distinguish the person from others. We call such piece of information a \textit{factoid}, which is modeled as a triplet consists of \textit{user identity}, \textit{predicate}, and an \textit{object}. By embedding these factoids, we learn the user identity representations and link two user identities if they are close to each other in the user embedding space. Our algorithm is carefully designed such that a piece of information, i.e. a factoid is ``translated'' into a motion in the user embedding space to bring two user identities closer if they belong to the same person or further part if not. Extensive experiments conducted on three real-world OSNs shows that \textsf{Factoid Embedding} outperforms the state-of-the-art methods. 
	\end{abstract}


	\section{Introduction}
	
	\textbf{Motivation.} Increasingly, people are using multiple online social networks (OSNs) to meet their communication and relationship needs\footnote{www.si.umich.edu/news/more-adults-using-multiple-social-platforms-survey-finds}. The rise of users using multiple OSNs motivates researchers to study User Identity Linkage (UIL), the problem of linking user accounts from different OSNs belonging to the same person. Tackling UIL is imperative to many applications, particularly user profiling and recommender systems.
	
	\textbf{User Identity Linkage Problem.} The UIL problem has been widely studied and is usually formulated as a classification problem, i.e. to predict whether a pair of user identities from different OSNs belong to the same real person \cite{DBLP:journals/sigkdd/ShuWTZL16}. There are many supervised and semi-supervised methods proposed to address UIL but they could not perform well when there is a \textit{scarcity of labeled data}. One possible way to obtain labeled matching user accounts is to recruit users to manually identify them. Such an approach is very costly and time consuming. In this research, we therefore aim  to solve the UIL problem using an unsupervised approach.
	
	\textbf{User Identity Linkage Problem in Unsupervised Setting.} We formulate the UIL problem in unsupervised setting as follows. Let $u$ be a user identity in an OSN which belongs to a real person $p$. Let $\mathbf{o}_u=[\mathbf{o}_{u,1}, ..., \mathbf{o}_{u,d}]$ denote a set of data objects associated with $u$. These objects include username, screen name, profile image, profile description, posts, etc.. We denote an OSN as $\mathcal{G}=(\mathcal{U}, \mathcal{O}, \mathcal{E})$, where $\mathcal{U}=\{u_1, ..., u_N\}$ is the set of user identities, $\mathcal{O}=\{\mathbf{o}_{u_1}, ..., \mathbf{o}_{u_N}\}$ is the set of corresponding data objects, and $\mathcal{E} \subseteq \mathcal{U} \times \mathcal{U}$ is the set of links in the network. Thus, given two OSNs $\mathcal{G}^s=(\mathcal{U}^s, \mathcal{O}^s, \mathcal{E}^s)$ (source) and $\mathcal{G}^t=(\mathcal{U}^t, \mathcal{O}^t, \mathcal{E}^t)$ (target), without any known matched user pairs between $\mathcal{G}^s$ and $\mathcal{G}^t$, the objective is to return a user $u^t$ in target OSN, for every user $u^s$ in the source OSN, such that the user pair $(u^s,u^t)$ most likely belongs to the same real person $p$. %Although the above formulation is for the two-OSNs setting, it can be easily extended to multiple OSNs. For simplicity, we also assume that each user does not have multiple accounts in one OSN. 

	\begin{figure*}[htb]
		\centering
		\includegraphics[width=0.95\textwidth]{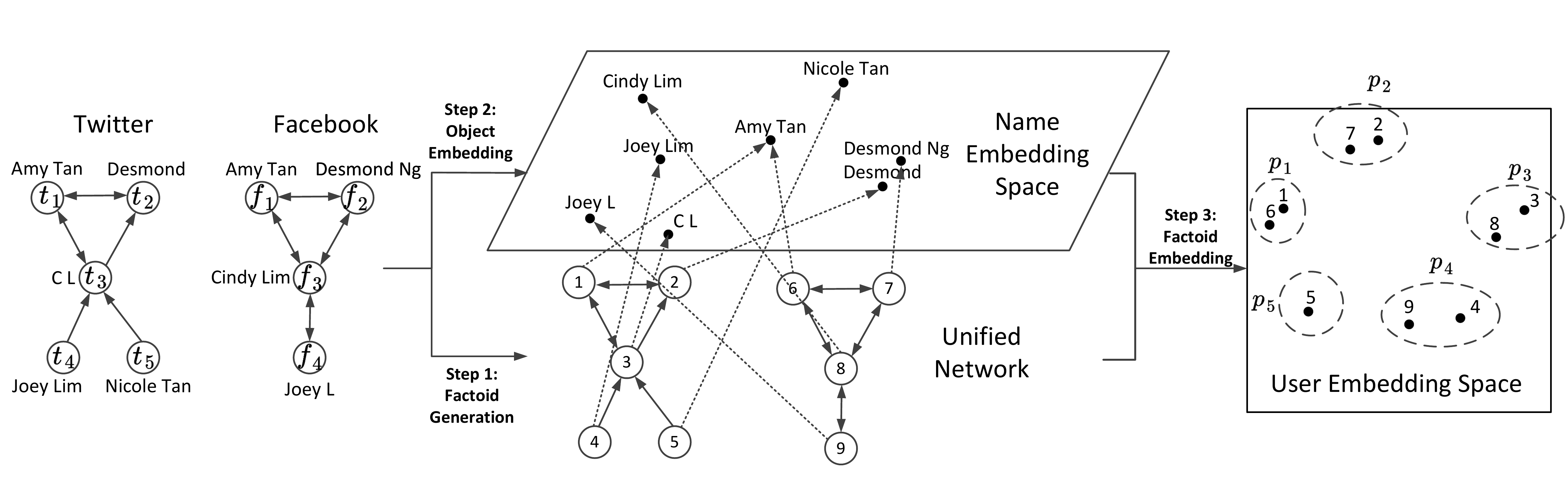}\\
		\caption{Framework of Factoid Embedding}
		\label{fig:workflow}
	\end{figure*}
	
	\textbf{Research Objectives and Contributions.} In this paper, we propose \textsf{Factoid Embedding}, a novel framework that links user identities across multiple OSNs through the use of a network embedding approach. The key idea behind \textsf{Factoid Embedding} is that despite the heterogeneity in information from multiple OSNs, each piece of information about a user identity describes the person who owns it, and thus help to distinguish the person from others. The more information we have, the closer we get to know about the real person. Specifically, we model each piece of information as a factoid, which is a triplet consisting of \textit{user identity}, \textit{predicate} and an \textit{object} or another \textit{user identity} (as shown in Table \ref{tbl:factoid}). Embedding these factoids provides the unifying structure to represent the heterogeneous information and data types. Figure \ref{fig:workflow} shows the framework of \textsf{Factoid Embedding}. Firstly, we generate the factoids from information gathered from different OSNs. Next, we embed heterogeneous objects (e.g. names, texts, and images) into their respective embedding spaces (e.g., names will be embedded into the name embedding space, etc.) incorporating the similarity measures that contribute to matching user identities. Note that when embedding the heterogeneous objects, we leverage on external and prior knowledge such that if two objects are similar, their embedding vectors will be close to each other in the object embedding space. For example, the user names \textit{Desmond} and \textit{Desmond Ng} are similar and therefore, the embedding vectors of the two names will be close to each other in the name embedding space. Finally, leveraging the factoids' triplet structure, we project the various object embeddings into the user embedding space. Essentially, the vectors in the user embedding space represent the user identities, and through iterations of object embedding projections, the user identities that share many similar factoids will be ``pushed'' closer to one another in the user embedding space.

	Overall, this paper improves the state-of-the-art by making the following contributions:
	
	\begin{itemize}
		\item We propose a novel unsupervised method called \textsf{Factoid Embedding} to link user identities from different OSNs. Our proposed method is able to integrate heterogeneous information using the triplet structure of factoids and object embeddings. To the best of our knowledge, this is the first work which embeds heterogeneous information to address the UIL problem in an unsupervised manner.
		\item We conduct extensive experiments on three real-world OSNs, namely, Twitter, Facebook and Foursquare, to evaluate our proposed model. The experiment results show that \textsf{Factoid Embedding} outperforms the state-of-the-art methods. It can even outperform some of the existing supervised methods which are given labeled matching user pairs for training.
		%\item We provide a case study and mathematical analysis to explain the dynamics in user embedding space. We demonstrate how our algorithm is able to ``translate'' a piece of information, i.e. a factoid, into a motion in the user embedding space to bring two user identities closer or further part. This allows us to empirically explain which factoids contributed to the linkage of two user identities.  
	\end{itemize}

	\section{Proposed Solution}
	\label{sec:solution}
	
	We propose an integrated three-step solution which is illustrated in Figure \ref{fig:workflow}. In the first step, we use the information from $\mathcal{G}^s$ and $\mathcal{G}^t$ to generate a set of factoids. Next, we embed the heterogeneous data objects (e.g. names, text and images) into their respective embedding spaces. Finally, we learn and project the user identities  and their links into user embedding space using factoid embedding, a process where we integrate the object embeddings and factoids.
	
	\begin{table}
		\centering
		\caption{Examples of Generated Factoids}
		\small%\footnotesize
		\label{tbl:factoid}
		\begin{tabular}{|l|l|}
			\hline
			Twitter&        Facebook           \\ \hline
			\multirow{10}{*}{}$\langle$1, \textit{has\_name},  Amy Tan$\rangle$& \multirow{10}{*}{}$\langle$6, \textit{has\_name},  Amy Tan$\rangle$\\
			$\langle$2, \textit{has\_name},  Desmond$\rangle$&$\langle$7, \textit{has\_name},  Desmond Ng$\rangle$                   \\
			$\langle$3, \textit{has\_name},  C L$\rangle$& $\langle$8, \textit{has\_name},  Cindy Lim$\rangle$                  \\
			$\langle$4, \textit{has\_name},  Joey Lim$\rangle$&$\langle$9, \textit{has\_name},  Joey L$\rangle$                   \\
			$\langle$5, \textit{has\_name},  Nicole Tan$\rangle$&$\langle$6, \textit{follows}, 7$\rangle$                \\ 
			$\langle$1, \textit{follows}, 2$\rangle$&$\langle$7, \textit{follows}, 6$\rangle$\\ 
			$\langle$2, \textit{follows}, 1$\rangle$&$\langle$6, \textit{follows}, 8$\rangle$\\
			$\langle$1, \textit{follows}, 3$\rangle$&$\langle$8, \textit{follows}, 6$\rangle$\\				
			$\langle$3, \textit{follows}, 1$\rangle$&$\langle$7, \textit{follows}, 8$\rangle$\\
			$\langle$3, \textit{follows}, 2$\rangle$&$\langle$8, \textit{follows}, 7$\rangle$\\
			$\langle$4, \textit{follows}, 3$\rangle$&$\langle$8, \textit{follows}, 9$\rangle$\\
			$\langle$5, \textit{follows}, 3$\rangle$&$\langle$9, \textit{follows}, 8$\rangle$\\
			
			\hline
		\end{tabular}
	\end{table}
	
	\subsection{Factoid Generation}
	\label{ssec:factoid}
	To integrate the heterogeneous user attribute/content objects and their user-user link information in $\mathcal{G}^s$ and $\mathcal{G}^t$, we first combine and represent the information in an unified network. In this unified network, every user identity $u_i$  (from source or target network) is represented as a new user node with a unique ID and every data object is represented as a data node (as illustrated in the step 1 in Figure \ref{fig:workflow}). We then represent a user-object association and a user-user link as an {\em user-object factoid} and an {\em user-user factoid} respectively. A user-object factoid $\langle u_i, pred, o \rangle$ has $pred$ denoting the associated attribute predicate, and $o$ denoting a data object. Each user-object factoid provides us a description about $u_i$. For example in Figure \ref{fig:workflow}, factoid $\langle$1, \textit{has\_name},  Amy Tan$\rangle$ conveys the information that the $u_1$ has name ``Amy Tan''. Next, we use another set of predicates to represent user-user links. For example, for Twitter, an user identity may ``follows'' another user identity. As such, we represent $follows$ as a predicate and let $\langle u_i, follows, u_j \rangle$ denote a user-user factoid with the predicate ``follows''. For instance, factoid $\langle$1, \textit{follows}, 3$\rangle$ tells us $u_1$ follows $u_3$. Table \ref{tbl:factoid} presents all the factoids generated from the two OSNs in Figure \ref{fig:workflow}. In the following, we shall elaborate the embeddings of objects followed by that of user-object and user-user factoids.

	\subsection{Object Embedding}
	Although the factoids generated in the previous step is able to represent the different information types in a unified network, it still has to address the issue of comparing data objects of attributes used for linking user identities. For example, the factoids in row 2 of Table \ref{tbl:factoid} do not explicitly tell us that ``Desmond'' and ``Desmond Ng'' are similar names. Instead, it only tell us that they are non-identical. Therefore, in this step we embed these heterogeneous objects taking advantage of similarity knowledge about the objects. For example, suppose two user identities sharing similar attribute objects are more likely to belong to the same person. We will then embed the objects such that similar objects are closer in the object embedding space (see step two in Figure \ref{fig:workflow}, where similar names are closer in the name embedding space).  
	
	We first let $O_{pred}$ denote all the data objects for certain predicate $pred$, i.e. $O_{pred} = \{o \in \mathcal{F}_{pred} \}$. For instance in Figure \ref{fig:workflow}, $O_{has\_name}$ = \{``Amy Tan'', ``Desmond'', ``C L'', ``Joey Lim'', ``Nicole Tan'', ``Desmond Ng'', ``Cindy Lim'', ``Joey L''\}. For each user-object predicate $pred$, we construct a similarity matrix $S^{pred}$ in which each element $S^{pred}_{i,j} \in [-1, 1]$ measures the similarity between two objects $o_i, o_j \in O_{pred}$. $S^{pred}_{i,j} =1$ when $o_i$ and $o_j$ are identical , and $=-1$ when $o_i$ and $o_j$ are completely different. There are a few ways to measure similarities between objects. For example, Jaro-Winkler distance \cite{DBLP:conf/ijcai/CohenRF03} has been used to measure the similarity between two names, and deep learning techniques can help us measure how similar two profile images are. In the experiment section, we will elaborate the similarities between different types of data objects in more details.
	
	For each data object $o$, we define $\mathbf{v}_o$ to be the embedding vector of $o$. To learn object embeddings, we define the objective function in Equation \ref{eq:error1}. This function aims to keep the embedding vectors of similar data objects to be close to each other in the object embedding space.
	
	\begin{equation}
	\label{eq:error1}
	error_{pred} = \sum_{i,j} (\mathbf{v}_{o_i}^{\top} \mathbf{v}_{o_j} - S^{pred}_{i,j})^2
	\end{equation}
	where $\{\mathbf{v}_{o}\}_{o \in O_{pred}}$ are the object embedding vectors, and $S^{pred}$ is the given similarity matrix. We learn $\{\mathbf{v}_{o}\}_{o \in O_{pred}}$ by minimizing $error_{pred}$.
	
	Ideally, $\{\mathbf{v}_{o}\}_{o \in O_{pred}}$ would preserve all the information in the similarity matrix $S^{pred}$ leading to  $error_{pred}$ = 0. More importantly, because $S^{pred}_{i,i}=1$, $\mathbf{v}_{o_i}^{\top} \mathbf{v}_{o_i} = ||\mathbf{v}_{o_i}||_2^2$ will be close to 1, i.e. all the embedding vectors $\{\mathbf{v}_{o}\}$ are near to the surface of a unit hypersphere. It means $\cos(\mathbf{v}_{o_i}, \mathbf{v}_{o_j}) \approx \mathbf{v}_{o_i}^{\top} \mathbf{v}_{o_j} \approx S^{pred}_{i,j}$. Therefore, if $o_i$ and $o_j$ are similar, $\mathbf{v}_{o_i}$ and $\mathbf{v}_{o_i}$ will be close to each other in the embedding space. Figure \ref{fig:workflow} illustrates how the names are embedded in the name embedding space. We can see that ``Desmond Ng'' is close to ``Desmond'', but far from ``C L'' in the embedding space.
	
	In practice, the similarity matrix $S^{pred}$ may be huge, i.e. $O((|\mathcal{U}^s| + |\mathcal{U}^t|)^2)$. In order to speed up learning, we can just focus on the similar object pairs. By employing blocking techniques such as inverted index and Locality-Sensitive Hashing (LSH) we can build a sparse similarity matrix $S^{pred}$. Afterwards, stochastic gradient descent is applied to minimize $error_{pred}$.\nop{ More details will be discussed in the experiment part.}

	\subsection{Factoid Embedding}
	In this step, we learn user identities' latent representations by embedding the generated user-object and user-user  factoids. 
	
	We let $\mathcal{U}^a$ denote the set of all user identities in the unified network (i.e. $\mathcal{U}^a = \mathcal{U}^s \cup \mathcal{U}^t$) and $\mathcal{F}_{pred}$ denote the set of factoids with a predicate $pred$, e.g. $\mathcal{F}_{follows} = \{\langle u_i, follows, u_j \rangle\}$, $\mathcal{F}_{has\_name} = \{\langle u_i, has\_name, o \rangle\}$. Suppose we have $d$ types of user-object predicates, i.e. $\{pred_1, ..., pred_d\}$.
	
	For each user-object factoid in $\mathcal{F}_{pred}$,  we define its probability as follows.
	
	\begin{equation}\label{eq:prob_pred}
	prob(u_i, pred, o) = \frac{exp(\mathbf{v}_{u_i}^{\top} \cdot \phi_{pred}(\mathbf{v}_o))}{\sum_{u' \in \mathcal{U}^a}exp(\mathbf{v}_{u'}^{\top} \cdot \phi_{pred}(\mathbf{v}_o))}
	\end{equation}
	where $\mathbf{v}_{u_i}$ is the embedding vector of user identity $u_i$, $\mathbf{v}_o$ is the embedding vector of data object $o$, and $\phi_{pred}$ is a projection function which maps $\mathbf{v}_o$ to the user embedding space. Note that we have learned $\mathbf{v}_o$ in the object embedding step. Particularly, we impose such a constraint on $\phi_{pred}$ that it is a Lipschitz continuous function, i.e. there is a a constant $C$ such that $|\phi_{pred}(\mathbf{x}) - \phi_{pred}(\mathbf{y})| < C\cdot|\mathbf{x} - \mathbf{y}|$ for any $\mathbf{x}$ and $\mathbf{y}$ in the space. In other words, if two objects are similar i.e. $\mathbf{v}_{o_i} \approx \mathbf{v}_{o_j}$ then their projections will be close to each other i.e. $\phi_{pred}(\mathbf{v}_{o_i}) \approx \phi_{pred}(\mathbf{v}_{o_j})$.\nop{We will show in Section \ref{ssec:dynamics} why this constraint is important for our solution.} In this work, we set $\phi_{pred}(\mathbf{v}_o)$ as a linear function, i.e. $\phi_{pred}(\mathbf{v}_o)=\mathbf{W}_{pred} \cdot \mathbf{v}_o + \mathbf{b}_{pred}$, where $\mathbf{W}_{pred}$ and $\mathbf{b}_{pred}$ are unknown parameters, and $\mathbf{W}_{pred}$'s norm $||\mathbf{W}_{pred}||$ is limited. We leave other non-linear choices of $\phi_{pred}(\mathbf{v}_o)$ for future work. Given all the user-object factoids in  $\mathcal{F}_{pred}$, i.e. $\{\langle u_i, pred, o \rangle\}$, we define the following objective function. 
	
	\begin{equation}\label{eq:obj_pred}
	f(\mathcal{F}_{pred}) = \sum_{\langle u_i, pred, o \rangle \in \mathcal{F}_{pred}} \log (prob(u_i, pred, o))
	\end{equation}
	
	Similarly, for each user-user factoid in $\mathcal{F}_{follows}$, we define its probability as follows.
	
	\begin{equation}\label{eq:prob_follows}
	prob(u_i, follows, u_j) = \frac{exp(\mathbf{v}_{u_i}^{\top} \cdot \phi_{follows}(\mathbf{v}_{u_j}))}{\sum_{u' \in \mathcal{U}^a}exp(\mathbf{v}_{u'}^{\top} \cdot \phi_{follows}(\mathbf{v}_{u_j}))}
	\end{equation}
	We set $\phi_{follows}(\mathbf{v}_u) = \mathbf{W}_{follows} \cdot \mathbf{v}_u + \mathbf{b}_{follows}$, where $\mathbf{W}_{follows}$'s norm $||\mathbf{W}_{pred}||$ is limited. Given all the user-user factoids in  $\mathcal{F}_{follows}$, i.e. $\{\langle u_i, follows, u_j \rangle\}$,we define the following objective function. 
	
	\begin{equation}\label{eq:obj_follows}
	f(\mathcal{F}_{follows}) = \sum_{\langle u_i, follows, u_j \rangle \in \mathcal{F}_{follows}} \log (prob(u_i, follows, u_j))
	\end{equation}
	
	We learn user embedding vectors $\{\mathbf{v}_u\}_{u \in \mathcal{U}^a}$ by solving the following multi-objective optimization problem.
	
	\begin{equation}
	\label{eq:opt}
	\max_{\{\mathbf{v}_u\}}(f(\mathcal{F}_{follows}), f(\mathcal{F}_{pred_1}), ..., f(\mathcal{F}_{pred_d}))
	\end{equation}
	Once we learned $\{\mathbf{v}_u\}_{u \in \mathcal{U}^a}$, we link user identities from different OSNs by simply comparing the distance between their embedding vectors. For example, in the user embedding space in Figure \ref{fig:workflow}, as user ID 1's nearest neighbor is user ID 6, we link them as the same person.
	
	\subsection{Optimization}

	To solve the multi-objective optimization in Equation \ref{eq:opt}, we optimize $f(\mathcal{F}_{follows}), f(\mathcal{F}_{pred_1}), \dotsc, f(\mathcal{F}_{pred_d})$ in turn. 
	
	As optimizing each objective function $f(\mathcal{F})$ is computationally expensive, we adopt the approach of negative sampling proposed in \cite{DBLP:conf/nips/MikolovSCCD13}. Particularly, for each factoid $\langle u_i, \cdot, \cdot \rangle$, $K$ ``fake'' factoids are introduced, i.e. $\{\langle u_k, \cdot, \cdot \rangle\}$, where $u_k$ are sampled from some noise distribution $P(u)$. More specifically, for a user-user factoid $\langle u_i, follows, u_j \rangle \in \mathcal{F}_{follows}$, we specifies the following objective function for it:
	\begin{equation}
	\label{eq:follow}
	\begin{split}
	f(u_i, follows, u_j) &= \log \sigma(\mathbf{v}_{u_i}^{\top} \cdot \phi_{follows}(\mathbf{v}_{u_j})) +\\
	&\sum_{k=1}^K E_{u_k \sim P_1(u)} [\log\sigma(-\mathbf{v}_{u_k}^{\top} \cdot \phi_{follows}(\mathbf{v}_{u_j}))]
	\end{split}
	\end{equation}
	where $\sigma(x)=\frac{1}{1 + \exp(-x)}$ is the sigmoid function. The first term models the observed factoid, and second term models the ``fake'' factoids and $K$ is the number of ``fake'' factoids. We set $P_1(u) \propto d_u^{3/4}$ as proposed in \cite{DBLP:conf/nips/MikolovSCCD13}, where $d$ is the out-degree of $u$ in the unified network. For a user-object factoid $\langle u_i, pred, o \rangle \in \mathcal{F}_{pred}$, its objective function is as follows.
	\begin{equation}
	\label{eq:attribute}
	\begin{split}
	f(u_i, pred, o) &= \log \sigma(\mathbf{v}_{u_i}^{\top} \cdot \phi_{pred}(\mathbf{v}_{o})) +\\
	&\sum_{k=1}^K E_{u_k \sim P_2(u)} [\log\sigma(-\mathbf{v}_{u_k}^{\top} \cdot \phi_{pred}(\mathbf{v}_{o}))]
	\end{split}
	\end{equation}
	And we set $P_2(u)$ as a uniform distribution over $\mathcal{U}^a$.

	Then stochastic gradient descent is used to optimize Equations \ref{eq:follow} and \ref{eq:attribute}. Algorithm \ref{alg:fe} gives an overview of \textsf{Factoid Embedding}. Suppose object embeddings and user embeddings have the same dimension $m$. The time complexity for each update operation in Algorithm \ref{alg:fe} is $O(K \cdot m^2)$. So the time complexity for Algorithm \ref{alg:fe} goes through all the user-object factoids and all the user-user factoids once are $O((|\mathcal{U}^s| + |\mathcal{U}^t|) \cdot d \cdot K \cdot m^2)$ and $O((|\mathcal{E}^s| + |\mathcal{E}^t|) \cdot K \cdot m^2)$ respectively.

	When we optimize $f(u_i, pred, o)$, we actually push $\mathbf{v}_{u_i}$ in the direction of $\phi_{pred}(\mathbf{v}_{o})$. It means that, user identities who share similar objects will be pushed towards each other. (There is also a similar effect for $f(u_i, follows, u_j)$.) This explains why \textsf{Factoid Embedding} is able to push similar user identities close to each other in the user embedding space.

	\begin{algorithm}[tb]
		\caption{Factoid Embedding}
		\label{alg:fe}
		\KwIn{$\mathcal{F}_{follows}, \mathcal{F}_{pred_1}, \dotsc, \mathcal{F}_{pred_d}$: the factoids with different predicates.}
		\KwIn{$\{\mathbf{v}_o\}_{o \in O_{pred_1}}, \dotsc, \{\mathbf{v}_o\}_{o \in O_{pred_d}}$: the object embeddings for user-object predicates: $pred_1, \dotsc, pred_d$.}
		\KwOut{$\{\mathbf{v}_u\}_{u \in \mathcal{U}^a}$.}
		
		Initialize $\{\mathbf{v}_u\}_{u \in \mathcal{U}^a}$, $\mathbf{W}_{follows}$, $\mathbf{b}_{follows}$, $\{\mathbf{W}_{pred_i}\}_{i=1}^d$, $\{\mathbf{b}_{pred_i}\}_{i=1}^d$\;
		\Repeat{convergence or reach maximum \# of iterations}{
			\For{$pred \in \{pred_1, \dotsc, pred_d\}$}{
				
				sample a batch of user-object factoids $\mathcal{F}^{\mathcal{B}}_{pred}$ from $\mathcal{F}_{pred}$;\\
				
				\For{$\langle u_i, pred, o \rangle \in \mathcal{F}^{\mathcal{B}}_{pred}$}{
					sample $K$ ``fake'' factoids $\{\langle u_k, pred, o \rangle\}_{k=1}^K$;\\
					update $\mathbf{v}_u$ according to $\frac{\partial{f(u_i, pred, o)}}{\partial{\mathbf{v}_{u}}}$;\\
				}
				
				update $\mathbf{W}_{pred}$ and $\mathbf{b}_{pred}$ (once for a certain \# of iterations);\\
				
			}
			
			sample a batch of user-user factoids $\mathcal{F}^{\mathcal{B}}_{follows}$ from $\mathcal{F}_{follows}$;\\
			
			\For{$\langle u_i, follows, u_j \rangle \in \mathcal{F}^{\mathcal{B}}_{follows}$}{
				sample $K$ ``fake'' factoids $\{\langle u_k, follows, u_j \rangle\}_{k=1}^K$;\\
				update $\mathbf{v}_u$ according to $\frac{\partial{f(u_i, follows, u_j)}}{\partial{\mathbf{v}_{u}}}$;\\
			}
			update $\mathbf{W}_{follows}$ and $\mathbf{b}_{follows}$ (once for a certain \# of iterations);\\
		}

		\Return{$\{\mathbf{v}_u\}_{u \in \mathcal{U}^a}$.}
		
	\end{algorithm}

	\section{Experiment}
	\label{sec:exp}

	\subsection{Data Collection}
	
	\nop{
		\begin{table*}[htb]
			\small
			\centering
			\caption{Datasets.}
			\label{tbl:data}
			\begin{tabular}{|c|C{2.5cm}|C{2.5cm}|C{1.5cm}|C{1.5cm}|}
				\hline
				Dataset& \multicolumn{2}{c|}{Facebook-Twitter}& \multicolumn{2}{c|}{Flickr-LiveJournal}  \\ \hline  \hline
				Network&  Facebook         &    Twitter    &  Flickr         &    LiveJournal          \\ \hline
				\# Users&      17,359     &    20,024       &      1,854    &   1,819     \\ \hline
				\# Links &      224,762     &      165,406    &      7,376    &      5,105      \\ \hline
				Available Information& \multicolumn{2}{c|}{username,screen name, profile image, network} & \multicolumn{2}{c|}{username, network}  \\ \hline
				\# Ground truth matching pairs& \multicolumn{2}{c|}{3,739} & \multicolumn{2}{c|}{52}  \\ \hline
				
			\end{tabular}
		\end{table*}
	}

	\begin{table*}[htb]
		\small
		\centering
		\caption{Datasets.}
		\label{tbl:data}
		%\vspace{-1mm}
		\begin{tabular}{|c|C{2.5cm}|C{2.5cm}|C{2.5cm}|C{2.5cm}|}
			\hline
			Dataset& \multicolumn{2}{c|}{Facebook-Twitter}& \multicolumn{2}{c|}{Foursquare-Twitter}  \\ \hline  \hline
			Network&  Facebook         &    Twitter    &  Foursquare         &    Twitter          \\ \hline
			\# Users&      17,359     &    20,024       &      21,668     &    25,772     \\ \hline
			\# Links &      224,762     &      165,406    &      312,740    &      405,590      \\ \hline
			Available Information& \multicolumn{2}{c|}{username,screen name, profile image, network} & \multicolumn{2}{c|}{screen name, profile image, network}  \\ \hline
			\# Ground truth matching pairs& \multicolumn{2}{c|}{1,998} & \multicolumn{2}{c|}{3,602}  \\ \hline
			
		\end{tabular}
	\end{table*}

	%We evaluate our proposed \textsf{Factoid Embedding}\footnote{Our code is available here https://github.com/linegroup/factoid\_embedding} by performing experiments on the following two real-world datasets.
	
	We evaluate our proposed \textsf{Factoid Embedding} using data sets from three popular OSNs, namely, Twitter, Facebook and Foursquare. We first gathered a set of Singapore-based Twitter users who declared Singapore as location in their user profiles. From the Singapore-based Twitter users, we retrieve a subset of Twitter users who declared their Facebook or Foursquare accounts in their short bio description as the ground truth. Table \ref{tbl:data} summarizes the statistics of our dataset.

	\nop{
		In total, we collected 1,998 Twitter-Facebook user identity pairs (known as \textit{TW-FB ground truth matching pairs}), and 3,602 Twitter-Foursquare user identity pairs (known as \textit{TW-FQ ground truth matching pairs}). 
		
		To simulate a real-world setting, where a user identity in the source OSN may not have its corresponding matching user identity in the target OSN, we expanded the datasets by adding Twitter, Facebook and Foursquare users who are connected to users in the \textit{TW-FB ground truth matching pairs} and \textit{TW-FQ ground truth matching pairs} sets. As such, although our evaluation is based on the \textit{TW-FB ground truth matching pairs} and \textit{TW-FB ground truth matching pairs}, the candidate user sets is much larger. Note that isolated users who do not have links to other users are removed from the data sets.

		After collecting the datasets, we extract the following user features using the OSNs' APIs. Table \ref{tbl:data} summarizes the statistics of our dataset.

		\begin{itemize}
			\item \textit{Username}: The username of the account (e.g., EunHye501).  
			%while username is not available for some Foursquare users.
			\item \textit{Screen name}: It is usually formed using the first and last name of the user (e.g., Eun Hye).
			\item \textit{Profile Image}: The thumbnail or image provided by the user.
			\item \textit{Network}: The relationship links between users.\nop{ Note that Facebook has an undirected friend relationships, while Twitter has directed following relationships.}  
		\end{itemize}
	}

	\subsection{Factoid Generation \& Object Embedding}
	The user-user and user-object factoids as described in Section \ref{ssec:factoid} are generated using the user information from the OSNs used in our experiment.
	
	We then calculate the similarity matrices for the data objects. We use Jaro-Winkler distance \cite{DBLP:conf/ijcai/CohenRF03} to measure username and screen name similarity\nop{\footnote{As Jaro-Winkler distance $dis$ is in the range of $[0,1]$, we calculate the final similarity $s=2*dis-1$ so that $s \in [-1,1]$.}}. \nop{In order to scale up the process, we build an 3-gram (i.e. 3 consecutive characters) inverted index for all the names. Only the similarities between names which share at least one 3-gram are calculated. In this way, we construct the following two sparse matrices: $S^{username}$ and $S^{screen\_name}$.} To measure the similarities between two profile images, we first use the deep learning model \textsf{VGG16} with weights pre-trained on ImageNet\footnote{https://keras.io/applications/\#vgg16} to extract a feature vector for each profile image. The cosine similarity between the two profile image feature vectors is then computed. \nop{An Locality-Sensitive Hashing (LSH) index on user profile images is build to speed up this process --- we only calculate the similarities between images in the same bucket. In this way, we construct the sparse matrix $S^{image}$.} Finally, we embed the data objects $\{\mathbf{v}_o\}_{o \in \mathcal{O}_{pred}}$ for each $pred$ (e.g. username) by using stochastic gradient descent to minimize $error_{pred}$ in Equation \ref{eq:error1}.
	
	\subsection{Evaluation Baselines and Metrics}
	
	\nop{	
		\begin{enumerate} 
			
			\item \textbf{Precision@K}. For each user account $a$ in the source social network, predict the $K$ most matching user accounts in the target social network, denoted by $U_p^{K}$. Precision@K is defined as follows.
			
			\[
			Precision@K = Avg_{a \in Q} Found(U_p^K, a)
			\]
			
			where $Found(U_p^K, a) = 1$ if the correct user account in the target social network is found in $U_p^K$, and $0$ otherwise.

			\item \textbf{Mean Reciprocal Rank (MRR)}.
			MRR is defined as follows.
			
			\[
			MRR = \frac{1}{|Q|} \sum_{a \in Q} \frac{1}{rank_a}
			\]
			
			where $rank_a$ denotes the rank position of the matching account for $a$.
			
		\end{enumerate}
	}

	%We compare our proposed \textsf{Factoid Embedding} with several state-of-the-art methods, which are divided into the following three categories: 
	
	\begin{enumerate}    
		\item \textbf{Supervised Methods}.
		
		\begin{itemize}
			
			\item \textbf{ULink} \cite{DBLP:conf/kdd/MuZLXWZ16} (S1): a supervised method which models the map from the observed data on the varied social platforms to the latent user space\nop{, such that the more similar the real users, the closer their profiles in the latent user space}. The node representations learned by Deepwalk\footnote{https://github.com/phanein/deepwalk}, concatenated with other object embedding vectors are used as user identity features. The code provided by the author is used for UIL.

			\item \textbf{Logistic Regression (LR)} (S2):\nop{ a classical binary classifier which is able to learn the importances of different features.} The following features are used: username similarity, screen name similarity, profile image similarity and the social status in network as defined in \cite{DBLP:conf/kdd/ZhangTYPY15}.

		\end{itemize}
		
		\item \textbf{Semi-Supervised Methods}.
		
		\begin{itemize}
			\item \textbf{COSNET} \cite{DBLP:conf/kdd/ZhangTYPY15} (SS1): an energy-based model which considers both local and global consistency among multiple networks. The candidate matching graph is generated based on profile-based features: username, screen name and profile image. The public code is used for UIL\footnote{https://aminer.org/cosnet}.
			
			\item \textbf{IONE} \cite{DBLP:conf/ijcai/LiuCLL16} (SS2): a network embedding based approach. Ground truth matching user identity pairs are needed to transfer the context of network structure from the source network to the target network. The original version of IONE uses network information only for UIL. For a fair comparison, we introduce more anchor links by linking user identities which share the same username, screen name or profile image.
			
			\item \textbf{Factoid Embedding* (FE*)} (SS3): Our proposed \textsf{Factoid Embedding} with labeled matching user identity pairs. Specifically, we adapt our solution to a semi-supervised version by merging the matching user identities into one node in the unified network. The merged user identities therefore share the same embedding vectors.
			
		\end{itemize}
		\item \textbf{Unsupervised Methods}.
		
		\begin{itemize}
			\item \textbf{Name} (U1): an unsupervised approach based on name similarity, which is reported as the most discriminative feature for UIL\cite{DBLP:conf/asunam/MalhotraTMKA12}. Here it can refer to username or screen name. We present whichever has the better performance.
			
			\item \textbf{CNL} \cite{DBLP:conf/icdm/GaoLLZPZ15} (U2): An unsupervised method which links users across different social networks by incorporating heterogeneous attributes and social features in a collective manner. The code provided by the author is used for UIL.
			
			\item \textbf{Factoid Embedding (FE)} (U3): Our proposed \textsf{Factoid Embedding} without any labeled matching user identity pairs.
		\end{itemize}

	\end{enumerate}

	For each ground truth matching \nop{user identity }pairs $(u^{s*}, u^{t*})$, we rank all the target users, i.e. $u^t \in \mathcal{G}^t$ according to $\cos(\mathbf{v}_{u^{s*}}, \mathbf{v}_{u^{t}})$. To quantitatively evaluate this ranking, we employ the following two metrics:

	\begin{itemize}
		\item \textbf{HitRate@K (HR@K)} in which a ranking is considered as correct if the matching user identity $u^{t*}$ is within the top $K$ candidates, i.e. $rank(u^{t*}) \leq K$.
		\item \textbf{Mean Reciprocal Rank (MRR)} is defined as follows.
		\[
		MRR = \frac{1}{n} \sum_{(u^{s*}, u^{t*})} \frac{1}{rank(u^{t*})}
		\]
		where $(u^{s*}, u^{t*})$ is a ground truth pair, and $n$ is the number of all the ground truth pairs.
	\end{itemize}

	\begin{table*}[htb]
		\small
		\centering
		\caption{Performance on Facebook-Twitter Dataset}
		\label{tbl:fb-tw}
		%\vspace{-1mm}
		\begin{tabular}{l|l|C{1.2cm}C{1.2cm}C{1.2cm}C{1.2cm}C{1.2cm}C{1.2cm}C{1.2cm}|C{1.2cm}}
			\hline \hline
			S/N&  Method&  HR@1     &   HR@2    &   HR@3 &   HR@4   & HR@5 &  HR@10  &   HR@30    &   MRR       \\ \hline
			S1& ULink &   \textbf{0.7071}&0.7285&0.7414&0.7471&0.7557&  0.7757 & 0.8042& 0.7102  \\ %\hline
			S2& LR &       0.5965   &0.6551 &0.6906  &0.7117 &0.7262& 0.7837  & 0.8098&0.6592      \\ \hline
			SS1& COSNET &0.6586  &0.7242  &0.7337 &0.7367&0.7382&  0.7417  & 0.7452& 0.6964      \\ %\hline
			SS2& IONE &  0.5605  &0.5695  &0.5725  &0.5730 &0.5750&  0.5805  & 0.6031&0.5698    \\ %\hline
			SS3& FE* &       0.6851  &\textbf{0.7322}  &\textbf{0.7567}  &\textbf{0.7747}  &0.7822 &  0.8098  &\textbf{0.8508}&\textbf{0.7297}     \\ \hline
			U1& Name &       0.5825  &0.6226  &0.6406  &0.6521  &0.6626&  0.6886 & 0.7232&0.6201    \\ %\hline
			U2& CNL &       0.5930    &0.6225	&0.6387	&0.6451	&0.6506&  0.6701  & 0.7327& 0.6284      \\ %\hline
			U3& FE&       0.6781  &0.7292  &0.7542  &0.7732  &\textbf{0.7827}& \textbf{0.8103}  &  0.8493& 0.7254  \\ \hline
			
		\end{tabular}
	\end{table*}

	\begin{table*}[htb]
		\small
		\centering
		\caption{Performance on Foursquare-Twitter Dataset}
		\label{tbl:fq-tw}
		%\vspace{-1mm}
		\begin{tabular}{l|l|C{1.2cm}C{1.2cm}C{1.2cm}C{1.2cm}C{1.2cm}C{1.2cm}C{1.2cm}|C{1.2cm}}
			\hline \hline
			S/N&  Method&  HR@1     &   HR@2    &   HR@3 &   HR@4   & HR@5 &  HR@10  &   HR@30    &   MRR       \\ \hline
			S1& ULink &  0.5464&  0.5843&  0.6032&  0.6232&  0.6399&  0.6766&  0.7397  &
			0.5915 \\ %\hline
			S2& LR &      0.5285  &0.5913  &0.6171  &0.6388  &0.6473&  0.6862 &0.7384&0.5882 \\ \hline
			SS1& COSNET &        0.5421 & 0.5905 & 0.6116 & 0.6238 &  0.6340&  0.6585 & 0.6693&0.5826   \\ %\hline
			SS2& IONE &   0.4081  &0.4158 &0.4225  &0.4269  &0.4297&  0.4408 & 0.4733 &0.4212      \\ %\hline
			SS3& FE* & \textbf{0.5541}  &\textbf{0.6021}  &\textbf{0.6293}  &\textbf{0.6440}  &\textbf{0.6546}&  \textbf{0.6979} & \textbf{0.7456}&\textbf{0.6029}    \\ \hline 
			U1& Name &     0.5227  &0.5730  &0.5980  &0.6154  &0.6332  &    0.6768  & 0.7293 &0.5741     \\ %\hline
			U2& CNL &       0.5283&  0.5786&  0.6050 &  0.6172&  0.6408&  0.6877&  0.7388&
			0.5853      \\ %\hline
			U3& FE&       0.5433  &0.5957  &0.6210  &0.6374  &0.6482&  0.6937 &0.7423&0.5944 \\ \hline
			
		\end{tabular}
	\end{table*}
	
	\subsection{Experimental Results}

	\textbf{Prediction Performance.} We randomly partition the ground truth matching user identity pairs into five groups and conduct five-fold cross-validation. Table \ref{tbl:fb-tw} presents the overall performance of the comparison methods on the Facebook-Twitter data set. It shows that, our proposed \textsf{Factoid Embedding} (SS3/U3) yields the best MRR result. Although \textsf{ULink} performs best on HR@1, \textsf{Factoid Embedding} outperforms it on both HR@K and MRR. The reason may be that, as a supervised approach \textsf{ULink} may link precisely the user identity pairs which can be represented by the training dataset. However, for the user identity pairs outside the labeled matching pairs, \textsf{ULink} may lose the ability to match them correctly. In contrast, by embedding factoids, \textsf{Factoid Embedding} is able to link such user identity pairs in an unsupervised manner. It explains why \textsf{ULink} has highest HR@1 but relatively low HR@30. It is a common problem for the supervised solutions for UIL because, as we mentioned in the introduction, the labeled dataset is quite small compared to the whole population. We also can observe that the \textsf{Factoid Embedding} outperforms the existing network embedding approach \textsf{IONE}(SS2), which makes use of the network information only. Interestingly, it can be seen that the performance of our unsupervised \textsf{Factoid Embedding} (U3) is very close to the semi-supervised version (SS3). One possible explanation is that SS3 just merges the matching pairs into one node, but does not learn from the labeled matching pairs like other supervised solutions e.g. \textsf{ULink}. Realizing that the performance of name similarity (U1) is relatively good, we think, by ``pushing'' similar user identities close to each other, U3 is able to ``merge'' these matching pairs in the user embedding space by itself. Thus the performances of U3 and SS3 are not significantly different. Table \ref{tbl:fq-tw} shows the results on the Foursquare-Twitter Dataset, which are consistent to that in Table \ref{tbl:fb-tw}. We can therefore conclude that our proposed \textsf{Factoid Embedding} performs best in both the unsupervised and supervised settings.

\begin{figure}[htb]
		\centering
		\begin{tabular}{cc}
			\includegraphics[width=1.5in]{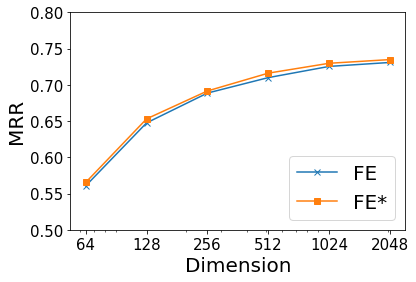}&\includegraphics[width=1.5in]{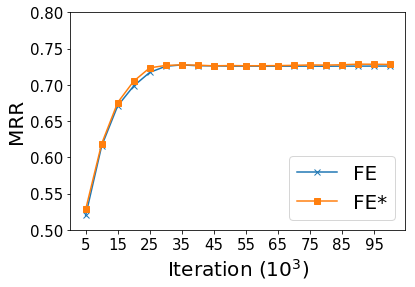}\\
			(a)&(b)
		\end{tabular}
		\caption{(a) Performance over Dimensions of User Embedding. (b) Performance over Iterations. }
		\label{fig:pa}
	\end{figure}

	\textbf{Parameter Analysis.} We investigate the performance w.r.t. the embedding dimension and the number of iterations on the Facebook-Twitter dataset. Figure \ref{fig:pa}(c) shows that the MRR performance of \textsf{Factoid Embedding} improves as the dimension increases. Figure \ref{fig:pa} (d) shows the MRR performances over different numbers of  iterations. We can see that the performance improves more significantly in the early iterations.

	%Due to the way we collected the ground truth data set, these set of users may be biased and not represent the whole user set. In order to fully evaluate our proposed solution, we employed human annotators to label the results returned by our solution for the non-ground truth users. Specifically, for a user identity $u^s \in \mathcal{G}^s$, we return the top user identity $u^t \in \mathcal{G}^t$ according to their distance in the user embedding space. The human annotators are given all the information in our dataset, also including the information outside our dataset -- they can go to the user's social network website homepage to explore and get a hint. When judging whether a given user identity pair belongs to the same person, these annotators have three options: ``match'', ``non-match'' and ``unknown''. The annotators are told they can choose ``match'' or ``non-match'' only when they have strong evidences about it, otherwise they should choose ``unknown''. It is worth noting that a ``non-match'' may be the result of wrong return of the solution, it also may because $u^s$'s corresponding user identity does not exist in our dataset, or $u^s$ does not have an account in the target OSN. Figure \ref{fig:pa} (d) presents the result. We can see that, compared to the baseline, \textsf{Factoid Embedding} match more user identity pairs, and at the same time, make less mistakes. Besides, it is surprising that even for human, for around 40\% user identity pairs they can not make a choice, which shows the intrinsic difficulty of UIL.

	\section{conclusion}
	\label{sec:conclude}
	
	In this paper, we proposed a novel framework \textsf{Factoid Embedding}, which adopts an unsupervised approach to cope with heterogeneity in user information and link users identities across multiple OSNs.  We evaluated \textsf{Factoid Embedding} using real-world datasets from three OSNs and benchmarked against the state-of-the-art UIL solutions. Our experimental results show that \textsf{Factoid Embedding} outperforms the state-of-the-art UIL solutions even in situations where the names of the user identities are dissimilar. 
	
	\nop{
		As there might be source OSN users who do not have their matching user identities in the target OSN, in the future we will investigate a post-processing step to determine if the matching target user identity exists. We also plan to explore more data object types, e.g. facial attributes from profile image.
	}

	\section*{Acknowledgment}
		This research is supported by the National Research Foundation, Prime Minister's Office, Singapore under its International Research Centres in Singapore Funding Initiative.

	\bibliographystyle{ieeetr}
	\bibliography{ref} 

\end{document}